\documentclass{aa}

\usepackage{natbib}
\bibpunct{(}{)}{;}{a}{}{,} 

\usepackage{graphicx}
\usepackage{amssymb}

\begin{document}

\title{TeV gamma rays from the blazar H~1426+428 and the 
diffuse extragalactic background radiation}

\titlerunning{TeV $\gamma$-rays from the blazar H~1426+428 and the 
diffuse extragalactic background radiation}

\author{
F.~Aharonian\inst{1},
A.~Akhperjanian\inst{7},
J.~Barrio\inst{3},
M.~Beilicke\inst{4},
K.~Bernl\"ohr\inst{1},
H.~B\"orst\inst{5},
H.~Bojahr\inst{6},
O.~Bolz\inst{1},
J.~Contreras\inst{2},
R.~Cornils\inst{4},
J.~Cortina\inst{2},
S.~Denninghoff\inst{2},
V.~Fonseca\inst{3},
M.~Girma\inst{1},
J.~Gonzalez\inst{3},
N.~G\"otting\inst{4},
G.~Heinzelmann\inst{4},
G.~Hermann\inst{1},
A.~Heusler\inst{1},
W.~Hofmann\inst{1},
D.~Horns\inst{1},
I.~Jung\inst{1},
R.~Kankanyan\inst{1},
M.~Kestel\inst{2},
J.~Kettler\inst{1},
A.~Kohnle\inst{1},
A.~Konopelko\inst{1},
H.~Kornmeyer\inst{2},
D.~Kranich\inst{2},
H.~Krawczynski\inst{1,9},
H.~Lampeitl\inst{1},
M.~Lopez\inst{3},
E.~Lorenz\inst{2},
F.~Lucarelli\inst{3},
N.~Magnussen\inst{6},
O.~Mang\inst{5},
H.~Meyer\inst{6},
R.~Mirzoyan\inst{2},
A.~Moralejo\inst{3},
E.~Ona\inst{3},
L.~Padilla\inst{3},
M.~Panter\inst{1},
R.~Plaga\inst{2},
A.~Plyasheshnikov\inst{1,8},
G.~P\"uhlhofer\inst{1},
G.~Rauterberg\inst{5},
A.~R\"ohring\inst{4},
W.~Rhode\inst{6},
J.~Robrade\inst{4},
G.~Rowell\inst{1},
V.~Sahakian\inst{7},
M.~Samorski\inst{5},
M.~Schilling\inst{5},
F.~Schr\"oder\inst{6},
I.~Sevilla\inst{3},
M.~Siems\inst{5},
W.~Stamm\inst{5},
M.~Tluczykont\inst{4},
H.J.~V\"olk\inst{1},
C.~A.~Wiedner\inst{1},
W.~Wittek\inst{2}}

\institute{Max-Planck-Institut f\"ur Kernphysik, Postfach 103980, D-69029 Heidelberg, Germany
\and Max-Planck-Institut f\"ur Physik, F\"ohringer Ring 6, D-80805 M\"unchen, Germany
\and Universidad Complutense, Facultad de Ciencias F\'{\i}sicas, Ciudad Universitaria, E-28040 Madrid, Spain
\and Universit\"at Hamburg, Institut f\"ur Experimentalphysik, Luruper Chaussee 149, D-22761 Hamburg, Germany
\and Universit\"at Kiel, Institut f\"ur Experimentelle und Angewandte Physik, Leibnizstra{\ss}e 15-19, D-24118 Kiel, Germany
\and Universit\"at Wuppertal, Fachbereich Physik, Gau{\ss}str.20, D-42097 Wuppertal, Germany
\and Yerevan Physics Institute, Alikhanian Br. 2, 375036 Yerevan, Armenia
\and On leave from Altai State University, Dimitrov Street 66, 656099 Barnaul, Russia
\and Now at Yale University, P.O. Box 208101, New Haven, CT 06520-8101, USA
}

\authorrunning{Aharonian et al.}

\date{Received / Accepted}

\offprints{\\G.~P\"uhlhofer,
\email{Gerd.Puehlhofer@mpi-hd.mpg.de}}

\abstract{The detection of TeV $\gamma$-rays from the blazar H\,1426+428 
at an integral flux level of 
$(4 \pm 2_{\mathrm{stat}} \pm 1_{\mathrm{syst}}) \times 10^{-12}\mathrm{erg}\,\mathrm{cm}^{-2}\mathrm{s}^{-1}$
above 1\,TeV with the HEGRA imaging atmospheric Cherenkov telescope system is reported.
H\,1426+428 is located at a redshift of $z$=0.129, which makes it the most distant source 
detected in TeV $\gamma$-rays so far.
The TeV radiation is expected to be strongly absorbed by the 
diffuse extragalactic background radiation (DEBRA).
The observed energy spectrum of TeV photons is in good agreement
with an intrinsic power law spectrum of the source $\propto E^{-1.9}$ corrected for DEBRA
absorption.
Statistical errors as well as uncertainties about the intrinsic source spectrum, however, do
not permit strong statements about the density of the DEBRA infrared photon field. 
\keywords{}
}

\maketitle

%

\section{Introduction}
\label{introduction}

Many nonthermal extragalactic objects representing different classes of 
AGNs are considered as potential sources of TeV photons. First of all
this concerns the BL\,Lac population of blazars in general;
two nearby representatives of this class, 
Mkn\,421 and Mkn\,501 with redshifts of $z$=0.030 and $z$=0.034, respectively,
are firmly established as TeV $\gamma$-ray emitters.
Of special interest are the so-called ``extreme synchrotron blazars'', 
BL\,Lac objects with flat spectra of synchrotron emission and 
high X-ray to radio flux ratios \citep[e.g.][]{ExtremeBLLacsBeppoSaX2001}.  
The observations of two such objects, Mkn\,501 
\citep{PianMrk501BeppoSax1997} and 1ES\,2344+514 \citep{Giommi2344BeppoSaX}, by BeppoSAX showed that 
in a flaring state the synchrotron peak of  ``extreme blazars'' can reach 100 keV.
Remarkably, both these objects are also reported as  
TeV blazars \citep[see e.g.][]{CataneseGammaRayAstronomy1999}. Most probably, this is not a 
mere coincidence; since the high synchrotron peak is an indicator of acceleration 
of electrons to ultrarelativistic energies, the ``extreme blazars'' are obvious 
candidates for TeV emission. Consequently,
some of these objects were intensively monitored with imaging atmospheric Cherenkov 
telescopes (IACT) over the last several years.
Possible detections of TeV signals have been claimed by different groups
for 1ES\,2344+514 \citep{CataneseWhipple2344}, 1ES\,1959+650 \citep{TelescopeArray1959ICRC}, 
and PKS\,2155-304 \citep{ChadwickDurham2155}.
Most recently, statistically significant TeV signals have 
been reported also for H\,1426+428, by the VERITAS \citep{Whipple1426Gamma2001}
and HEGRA \citep{Talk1426HEGRAICRC2001,TalkFelixICRC2001} collaborations.
The relatively large redshift ($z$=0.129) of H\,1426+428,
as well as the fact that it is the 
third 100 keV blazar ever found \citep{ExtremeBLLacsBeppoSaX2001}, make this object 
an extremely important target for future multiwavelength studies of 
the nonthermal processes of particle acceleration and radiation in AGN jets, as 
well as for probing the DEBRA in the near infrared region.    

In this letter we report the HEGRA observations of H\,1426+428 performed 
in 1999 and 2000, and briefly discuss some apparent  
astrophysical implications suggested by the obtained results.    
 

\section{Observations and results of analyses}
\label{observations}

For the following analysis,
14.5\,hrs of data from 1999 and 29.9\,hrs from 2000 were available,
taken with the HEGRA IACT system;
see Table \ref{datalog} for the dates of observations.
The median zenith angle was 17\degr, resulting in 
a $\gamma$-shower peak detection rate at an energy of $700\,\mbox{GeV}$
for \object{Crab}-like spectra \citep{HEGRAPerformance99}.

\begin{table}[b]
  \centering
  \begin{tabular}{ll} \hline \hline
  1999:                 & Feb. 23., Mar. 15., 16., 18.-22., Apr. 10.-12., 20.     \\ \hline
  2000:                 & Mar. 5.-7., 14., 28.-30., Apr. 3.-5., 26.-28., 30., \\
                        & May 1., 6.-8., 29., 30., Jun. 1.-5. \\   
  \hline \hline
  \end{tabular}
  \caption[]{Dates of observations. Each date refers to the respective following night,
  with typically one hour of observations.}
  \label{datalog}
\end{table}

\begin{table}[b]
  \centering
  \begin{tabular}{lll} \hline \hline
                                                & signal           & spectrum  \\ \hline
  stereo algorithm                              & \#3              & \#1                \\
  shape cut: mean scaled width $<$              & 1.1              & 1.2                \\
  angular cut: $\theta^{2} <$                   & 0.011\,deg$^2$   & 0.05\,deg$^{2}$    \\ \hline
  $N_{\mathrm{off}}$                            & 1779             & 6258               \\
  $\alpha=A_{\mathrm{on}}/A_{\mathrm{off}}$     & 0.143            & 0.262              \\
  $N_{\mathrm{on}}$                             & 360              & 1839               \\
  $N_{\mathrm{\gamma-candidates}}$              & 105.9            & 199.2              \\ 
  significance $\sigma$                         & 5.8              & 4.3                \\ \hline
  \hline
  \end{tabular}
  \caption[]{Cuts, event numbers, and significances for the signal search and the spectral analysis. 
  $A$: on resp. off source area.}
  \label{numbers}
\end{table}

Shower reconstruction and cuts were described e.g. in \citet{Mrk501PaperI}.
The methods applied in this analysis 
differ slightly, depending on the tasks which were performed. For the
search for new sources we applied tight cuts and stereo algorithm \#3 \citep{HofmannStereoTechniques}
to achieve high sensitivity. For the spectral
analysis of the detected source, somewhat relaxed cuts were used which provide a larger 
$\gamma$-acceptance; these cuts are less susceptible to systematic errors, and the spectral
reconstruction was tested in great detail using Monte Carlo studies and $\gamma$-ray data from
known TeV sources \citep{Mrk501PaperI,Mrk501TimeAveraged}. 
The background was estimated using a set of control regions
in the field of view (FOV); the setup provides seven times more background data
than on source data under identical acceptance conditions \citep[][]{HEGRACasAPaper}.
The optimum angular cut (for the signal search) was derived on the basis of
the angular resolution -- using $\gamma$-ray events from the Crab nebula -- and the background level.
Table~\ref{numbers} summarizes cut parameters, resulting event numbers, and significances
for the signal search and spectral analysis.

\begin{figure}
  \centering
  \includegraphics[width=5.5cm]{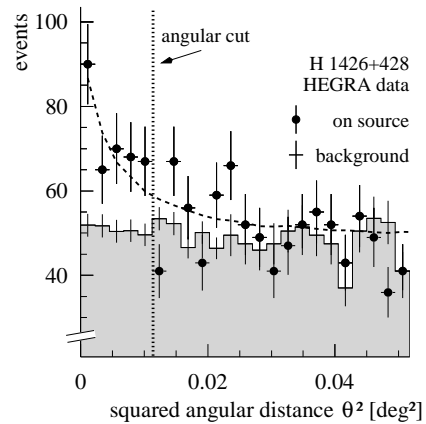}
  \caption[]{Dots: Number of events vs. the squared angular distance
  to the position of H\,1426+428.
  Shaded histogram: Background estimate; 
  up to $\theta^{2}=0.0225\,\mathrm{deg}^{2}$, data from 7 control regions are used.
  Therefore, the statistical error of the background estimate is much smaller than the
  error of the source distribution.
  The dashed line shows Crab excess events, measured at similar zenith angles, 
  scaled down to 6\%,
  and superimposed on a flat background.
  The vertical dotted line indicates the position of the optimum angular cut, based on Crab data. }
  \label{bl1426excess}
\end{figure}

\begin{figure}
  \centering
  \includegraphics[width=5.5cm]{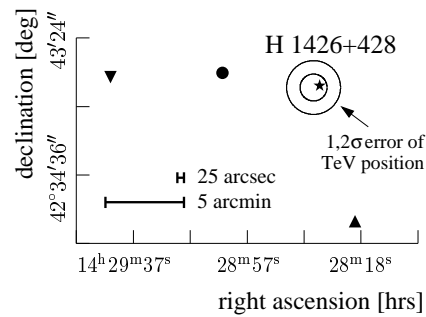}
  \caption[]{The sky area around H\,1426+428 in celestial coordinates (epoch 2000).
             The circles show the 1 and 2 $\sigma$ error 
             of the reconstructed TeV position (HEGRA data).
             The systematic pointing uncertainty of the HEGRA system is 25 arcsec.
             The positions of X-ray selected QSOs 
             ($\star$: H\,1426+428 (position from Hubble space telescope observations), 
	     $\blacktriangle$: CRSS J1428.3+4231, 
	     $\blacktriangledown$: CRSS J1429.7+4240) and galaxies
             ($\bullet$: the galaxy group CRSS J1429.1+4241) were obtained from NED. 
             }
  \label{bl1426position}
\end{figure}

Figure~\ref{bl1426excess} shows the event distribution, obtained from the signal search,
as a function of the squared angular distance to the source position. 
The excess significance, calculated according to \citet{LiMa}, amounts to $5.8\,\sigma$
($2.4\,\sigma$ in 1999, $5.3\,\sigma$ in 2000). 
While there is an indication that the source was in a higher state in May 2000, the total data
set is also statistically consistent with constant signal accumulation. 

The event distribution in the FOV was used to reconstruct the source position as seen in 
TeV $\gamma$-rays; for details of the procedure see \citet{HEGRAPointing} and \citet{HEGRACasAPaper}.
Figure~\ref{bl1426position} shows the sky area around H\,1426+428. The positions of 
X-ray selected QSOs and galaxies were obtained from the NASA/IPAC Extragalactic Database (NED).
The median angular resolution of the HEGRA system is approx. 0\fdg1; the given statistics allowed the
reconstruction of the center of TeV emission with a statistical error 
of 50\arcsec (1\,$\sigma$).
A confusion of the TeV source with other known X-ray sources is excluded. 
We note that the very prominent X-ray source GB\,1428+4217
is located at an angular distance of 41\farcm2 to H\,1426+428, far outside the area shown
in Fig.~\ref{bl1426position}.

\begin{figure}
  \centering
  \includegraphics[width=6cm]{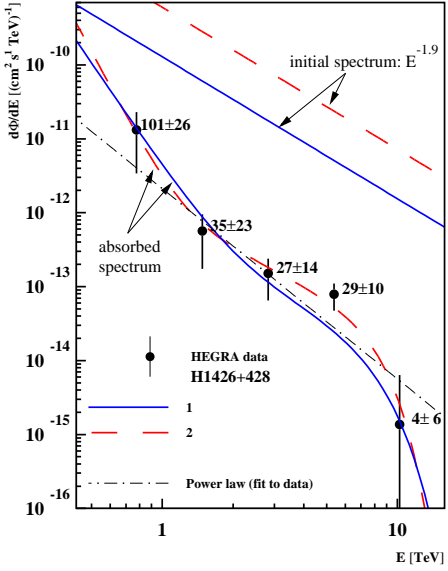}
  \caption[]{Differential energy spectrum of H\,1426+428 (dots); the numbers indicate the
  (background subtracted) event counts and their corresponding 
  statistical $1\sigma$-errors. A power
  law fit to the data is shown by the dashed-dotted line. 
  The solid and the dashed lines represent absorbed model spectra, obtained by an initial 
  power law spectrum $\propto E^{-1.9}$ (normalized to the data)
  combined with 
  different absorption models (see Fig.~\ref{panelirfield}).}
  \label{bl1426spectrum}
\end{figure}

The energy reconstruction, which has a single event resolution of $\Delta E/E = 20\%$,
and the spectral evaluation were described in detail in \citet{Mrk501PaperI,Mrk501TimeAveraged}.
The energy spectrum was derived from the raw, background subtracted photon count spectrum using a
so called effective area, which is 
a response function depending upon the reconstructed energy and zenith angle;
the effective area was adjusted regularly according to the varying detector conditions and
different system setups (3/4/5 telescopes included in the system).

Figure~\ref{bl1426spectrum} shows the reconstructed differential energy spectrum of H\,1426+428. 
To first order, the spectrum can be described by a pure power law
$\mathrm{d}\Phi/\mathrm{d}E = (2.0 \pm 1.3_{\mathrm{stat}} \pm 0.1_{\mathrm{syst}})
\times 10^{-12}(E/\mathrm{TeV})^{(-2.6 \pm 0.6_{\mathrm{stat}} \pm
0.1_{\mathrm{syst}})}\,\mathrm{ph}\,\mathrm{cm}^{-2}\,\mathrm{s}^{-1}\,\mathrm{TeV}^{-1}$
(the systematic errors do not include the 15\% error on the energy scale).
The spectral slope is similar to the one measured for 
the Crab nebula \citep{KonopelkoCrab}.  
The measured integral flux (derived using loose cuts) above 1\,TeV is 
$(1.7 \pm 0.5_{\mathrm{stat}} \pm 0.1_{\mathrm{syst}}) \times 10^{-12}\mathrm{ph}\,\mathrm{cm}^{-2}\,\mathrm{s}^{-1}$
which corresponds to 10\% of the Crab nebula flux. A comparison of the total photon 
count rates of H\,1426+428 and Crab nebula after tight cuts yields 6\%; within statistical errors,
both results are compatible.

%

\section{Discussion}
\label{implications}

\begin{figure}
  \centering
  \includegraphics[width=7cm]{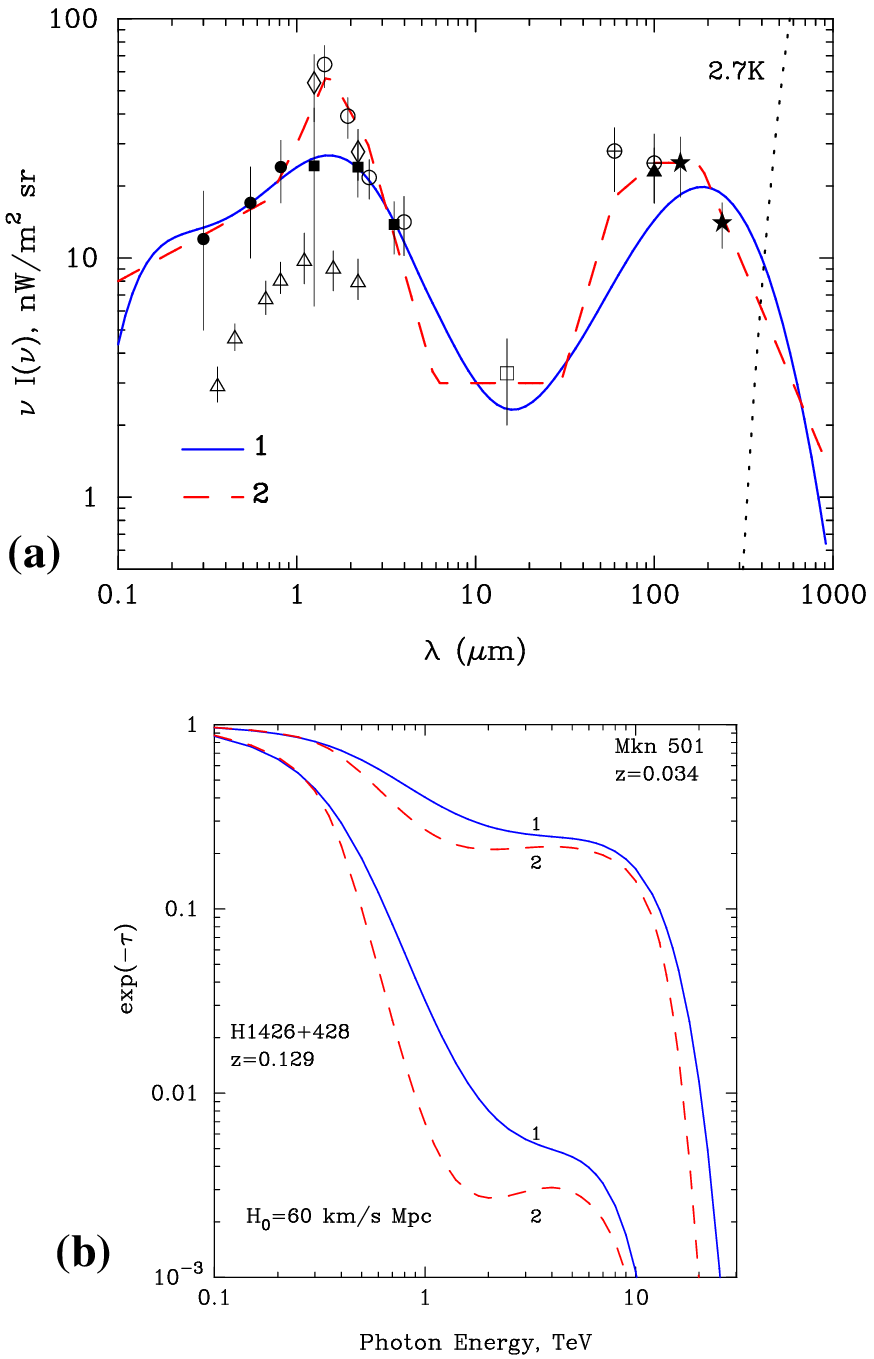}
  \caption[]{{\bf (a)} Diffuse extragalactic background radiation. The reported fluxes at 
  far infrared wavelengths above $60\,\mu \rm m$,   
  as well as at optical wavelengths below $1\,\mu \rm m$ 
  are taken from the recent review by \citet{IRHauser2001}.
  The lower flux limit 
  at $15\,\mu \rm m$ is derived from the ISOCAM source counts
  \citep{IRFranceschini2001}. The fluxes reported at NIR are shown by 
  filled squares \citep{IRWright2001}, open circles \citep{IRMatsumoto2000}, and 
  open diamonds \citep{IRCambresy2001}. Two model curves are shown by dashed and 
  solid lines, respectively.
  {\bf (b)} The spectrum modification factors $\exp[-\tau(E)]$
  calculated for the two DEBRA models for Mkn\,501 and H\,1426+428.}
\label{panelirfield}
\end{figure}

For any reasonable DEBRA model (see Fig.~\ref{panelirfield}\,a),
the TeV radiation from H\,1426+428 is expected to arrive with a  
drastically modified spectrum.
Reasons are the strong energy dependence of the 
mean free path of $\gamma$-rays in the intergalactic medium, $\Lambda(E)$,
and the fact that the optical depth 
$\tau=c\,z\,H_0^{-1}\,\Lambda^{-1}(E)$ significantly exceeds unity at all $\gamma$-ray energies
above 300\,GeV for H\,1426+428 with a redshift of $z$=0.129. 
The conclusion about strong absorption at all energies covered by HEGRA
does not depend on the large systematic uncertainties in the reported DEBRA fluxes.
This is demonstrated in 
Fig.~\ref{panelirfield}\,b where we show the intergalactic absorption factors
calculated for two reference DEBRA models. 
Model 1 is close to the recent calculations of 
\citet{IRPrimackHDGS2000} based on the so-called Kennicut 
initial mass function (IMF), and fits quite well the DEBRA fluxes  
reported recently by \citet{IRWright2001} at 1.25, 2.2 and 3.5 $\mu\mathrm{m}$
wavelengths. Model 2 is designed to match the NIR fluxes reported by 
\citet{IRMatsumoto2000} and \citet{IRCambresy2001}
who claim very high fluxes at wavelengths shorter than $2\,\mu \rm m$. 
Despite significant quantitative differences, the two 
DEBRA models result in similar absorption features in the TeV spectrum:  
both models predict a very strong steepening of the spectrum at energies below 1-2 TeV,
but an almost energy independent absorption at energies around several TeV 
(see Fig.~\ref{panelirfield}\,b). 
This effect is common for all 
realistic model curves describing the DEBRA in the near IR
\citep[e.g.][]{IRPrimackHDGS2000},
since they reflect the characteristic shape of the starlight 
spectrum which in the wavelength band between 1 and several microns
behaves as $\nu I(\nu) \propto \lambda^{\beta}$ with $\beta \sim -1$.
The corresponding number density of background photons 
$n(\epsilon) \propto \epsilon^{-1}$ results in a nearly constant optical depth 
for $\gamma$-rays at energies around several TeV \citep{Mrk501TimeAveraged}. 

The X-ray spectrum of H\,1426+428 which extends up to 100\,keV favours an intrinsic pure power law
spectrum in the relevant energy range, with an index of 1.9
(a detailed discussion is beyond the scope of this paper and will be given elsewhere).
The observed spectrum -- after absorption -- is therefore not expected to obey a power law.
The results presented in Fig.~\ref{bl1426spectrum} show that the HEGRA spectral points 
are indeed better described by an {\it intrinsic} power law spectrum  
{\it modified by intergalactic absorption}. 
However, a power law fit with an adjustable spectral index results in a reduced 
$\chi^{2}/\mathrm{dof}=4.4/3=1.47$, while 
absorbed spectra -- with a fixed source spectrum $\propto E^{-1.9}$ -- yield 
$\chi^{2}/\mathrm{dof}=3.7/4=0.9$ and $\chi^{2}/\mathrm{dof}=1.0/4=0.25$, 
for the DEBRA models 1 and 2 (see Fig.~\ref{panelirfield}), respectively.
Hence, more accurate spectral measurements, as well as 
detailed theoretical studies (concerning both the formation of the 
$\gamma$-ray spectra in the source and their subsequent 
deformation during the passage through the extragalactic photon fields),
are required.

Nevertheless, the sheer fact of detection of TeV $\gamma$-rays 
from H\,1426+428 should lead to a revision of the current 
conceptual view of TeV blazars, according to which the synchrotron (X-ray)
peak in the SED dominates over the inverse Compton (TeV) peak 
\citep[see e.g.][]{FossatiBlazars1998}.
Indeed, although the detected energy flux of $\gamma$-rays is only  
$\simeq 4 \times 10^{-12}\,\rm erg\,cm^{-2}s^{-1}$, corrected for intergalactic 
absorption this flux may well exceed $10^{-10}\,\rm erg\,cm^{-2}s^{-1}$. 
For comparison, the X-ray flux measured  
by BeppoSAX \citep{ExtremeBLLacsBeppoSaX2001} and 
ASCA (T. Takahashi, private communication) is well below $10^{-10}\,\rm erg\,cm^{-2}s^{-1}$. 
Since the corrected TeV luminosity seems to exceed the level anticipated 
from the current models of TeV blazars by far, this result may have a crucial impact on 
the further development of models for TeV blazars.
 
H\,1426+428 is the most distant source detected in TeV $\gamma$-rays so far.
The HEGRA collaboration intends to continue observing this source extensively in 2002
in the hope to decisively improve the statistics of the measurement.

\begin{acknowledgements}
The support of the German ministry for research and
technology (BMBF) and of the Spanish Research Council (CICYT) is gratefully
acknowledged. GR acknowledges receipt of a Humboldt fellowship.
We thank the Instituto de Astrof\'{\i}sica de Canarias
for the use of the site and for supplying excellent working conditions at
La Palma. 
\end{acknowledgements}

\bibliographystyle{apj}
\bibliography{letter}

\begin{thebibliography}{25}
\expandafter\ifx\csname natexlab\endcsname\relax\def\natexlab#1{#1}\fi

\bibitem[{Aharonian(2001)}]{TalkFelixICRC2001}
Aharonian, F.~A. 2001, in Proc. of the 27th ICRC, astro--ph/0112314

\bibitem[{Aharonian {et~al.}(1999{\natexlab{a}})Aharonian, Akhperjanian,
  Barrio, {et~al.}}]{Mrk501PaperI}
Aharonian, F.~A., Akhperjanian, A.~G., Barrio, J.~A., {et~al.}
  1999{\natexlab{a}}, A\&A, 342, 69

\bibitem[{Aharonian {et~al.}(1999{\natexlab{b}})Aharonian, Akhperjanian,
  Barrio, {et~al.}}]{Mrk501TimeAveraged}
---. 1999{\natexlab{b}}, A\&A, 349, 11

\bibitem[{Aharonian {et~al.}(2000)Aharonian, Akhperjanian, Barrio,
  {et~al.}}]{KonopelkoCrab}
---. 2000, ApJ, 539, 317

\bibitem[{Aharonian {et~al.}(2001)Aharonian, Akhperjanian, Barrio,
  {et~al.}}]{HEGRACasAPaper}
---. 2001, A\&A, 370, 112

\bibitem[{Cambresy {et~al.}(2001)Cambresy, Reach, Beichman, \&
  Jarrett}]{IRCambresy2001}
Cambresy, L., Reach, W.~T., Beichman, C.~A., \& Jarrett, T.~H. 2001, ApJ, 555,
  563

\bibitem[{Catanese {et~al.}(1998)Catanese, Akerlof, Badran,
  {et~al.}}]{CataneseWhipple2344}
Catanese, M., Akerlof, C.~W., Badran, H.~M., {et~al.} 1998, ApJ, 501, 616

\bibitem[{Catanese \& Weekes(1999)}]{CataneseGammaRayAstronomy1999}
Catanese, M. \& Weekes, T. 1999, PASP, 111, 1193

\bibitem[{Chadwick {et~al.}(1999)Chadwick, Lyons, McComb,
  {et~al.}}]{ChadwickDurham2155}
Chadwick, P.~M., Lyons, K., McComb, T. J.~L., {et~al.} 1999, ApJ, 513, 161

\bibitem[{Costamante {et~al.}(2001)Costamante, Ghisellini, Giommi,
  {et~al.}}]{ExtremeBLLacsBeppoSaX2001}
Costamante, L., Ghisellini, G., Giommi, P., {et~al.} 2001, A\&A, 371, 512

\bibitem[{Fossati {et~al.}(1998)Fossati, Maraschi, Celotti,
  {et~al.}}]{FossatiBlazars1998}
Fossati, G., Maraschi, L., Celotti, A., {et~al.} 1998, MNRAS, 299, 433

\bibitem[{Franceschini {et~al.}(2001)Franceschini, Aussel, Cesarsky,
  {et~al.}}]{IRFranceschini2001}
Franceschini, A., Aussel, H., Cesarsky, C.~J., {et~al.} 2001, A\&A, 378, 1

\bibitem[{Giommi {et~al.}(2000)Giommi, Padovani, \&
  Perlman}]{Giommi2344BeppoSaX}
Giommi, P., Padovani, P., \& Perlman, E. 2000, MNRAS, 317, 743

\bibitem[{G{\"o}tting {et~al.}(2001)}]{Talk1426HEGRAICRC2001}
G{\"o}tting, N. {et~al.} 2001, talk at the 27th ICRC, Hamburg

\bibitem[{Hauser \& Dwek(2001)}]{IRHauser2001}
Hauser, M. \& Dwek, E. 2001, ARA\&A, 39, 249

\bibitem[{Hofmann {et~al.}(1999)Hofmann, Jung, Konopelko,
  {et~al.}}]{HofmannStereoTechniques}
Hofmann, W., Jung, I., Konopelko, A., {et~al.} 1999, Astropart. Phys., 12, 135

\bibitem[{Horan \& the VERITAS~coll.(2001)}]{Whipple1426Gamma2001}
Horan, D. \& the VERITAS~coll. 2001, in GAMMA 2001: Gamma-Ray Astrophysics
  2001, AIP Conf Proc. 587, 324 ff.

\bibitem[{Konopelko {et~al.}(1999)Konopelko, Hemberger, Aharonian,
  {et~al.}}]{HEGRAPerformance99}
Konopelko, A., Hemberger, M., Aharonian, F., {et~al.} 1999, Astropart. Phys.,
  10, 275 ff.

\bibitem[{Li \& Ma(1983)}]{LiMa}
Li, T.-P. \& Ma, Y.-Q. 1983, ApJ, 272, 317

\bibitem[{Matsumoto(2000)}]{IRMatsumoto2000}
Matsumoto, T. 2000, Report No. 14, The Institute of Space and Astronautical
  Science, Kanagawa, Japan

\bibitem[{Nishiyama {et~al.}(1999)Nishiyama, Chamoto, Chikawa,
  {et~al.}}]{TelescopeArray1959ICRC}
Nishiyama, T., Chamoto, N., Chikawa, M., {et~al.} 1999, in Proc. of the 26th
  ICRC, Salt Lake City, vol. 3, 370 ff.

\bibitem[{Pian {et~al.}(1998)Pian, Vacanti, Tagliaferri,
  {et~al.}}]{PianMrk501BeppoSax1997}
Pian, E., Vacanti, G., Tagliaferri, G., {et~al.} 1998, ApJ, 492, L17

\bibitem[{Primack {et~al.}(2001)Primack, Somerville, Bullock, \&
  Devriendt}]{IRPrimackHDGS2000}
Primack, J.~R., Somerville, R.~S., Bullock, J.~S., \& Devriendt, J. E.~G. 2001,
  in High Energy Gamma-Ray Astronomy: Intern. Symp., AIP Conf Proc. 558, 463 ff

\bibitem[{P{\"u}hlhofer {et~al.}(1997)P{\"u}hlhofer, Daum, Hermann,
  {et~al.}}]{HEGRAPointing}
P{\"u}hlhofer, G., Daum, A., Hermann, G., {et~al.} 1997, Astropart. Phys., 8,
  101

\bibitem[{Wright \& Johnson(2001)}]{IRWright2001}
Wright, E.~L. \& Johnson, B.~D. 2001, ApJ, submitted

\end{thebibliography}

\end{document}